%% file: qqbZZ.tex
\documentclass[a4paper,11pt]{article}
\pdfoutput=1 
\usepackage{jheppub}  
\usepackage{graphicx,color}
\usepackage{amsmath}
\usepackage{autobreak}
\usepackage[normalem]{ulem}
\allowdisplaybreaks
\newcommand{\ben}{\begin{enumerate}}
\newcommand{\een}{\end{enumerate}}
\newcommand{\beq}{\begin{equation}}
\newcommand{\eeq}{\end{equation}}
\newcommand{\bal}{\begin{align}}
\newcommand{\eal}{\end{align}}

\newcommand{\bea}{\begin{eqnarray}}
\newcommand{\eea}{\end{eqnarray}}
\newcommand{\nn}{\nonumber\\}

\def\z#1{{\zeta_{#1}}}

\newcommand{\Ca}{C_A}
\newcommand{\Cf}{C_F}
\newcommand{\nf}{n_F}

\newcommand{\df}{{\rm d}}
\def\Dm1{{{\delta(1-z)}}}

\def\nf{{n^{}_{\! f}}}

\def\cf{{C^{}_F}}

\def\A#1{{A_{#1}}}

\def\g0#1DY{{g_{0#1}^{DY}}}

\def\gNB#1{{g_{#1}}}
\newcommand{\Lqr}{L_{qr}}
\newcommand{\Lfr}{L_{fr}}

\newcommand{\iW}{\omega^{-1}}
\def\LogmW1{{{\ln (1-\omega)}}}
\def\w{{\omega}}
\def\WbimW{{\frac{\omega}{(1-\omega)}}}
\def\LogomWtIMW{{\frac{\ln(1-\omega)}{(1-\omega)}}}

\def\LogtmWtIMW{{\frac{\ln (1-\omega)^2}{(1-\omega)}}}

\def\btzAIII{{\frac{A_3}{\beta_0^2}}}
\def\btzAII{{\frac{A_2}{\beta_0}}}
\def\btzAI{{A_1}}

\def\btzDII{{\frac{D_2}{\beta_0}}}
\def\btzDI{{D_1}}

\def\AAo{{\frac{A_1}{\beta_{0}}}}
\def\AAt{{\frac{A_2}{\beta_{0}^2}}}

\def\DDo{{\frac{D_1}{\beta_{0}}}}

\newcommand{\btoo}{{\bigg(\frac{\beta_{2}}{\beta_0^{3}}\bigg)}}

\newcommand{\overbar}[1]{\,\overline{\!{#1}}}
\newcommand{\Nbar}{\overbar{N}}
\newcommand{\gbar}{\overbar{g}}
\newcommand{\as}{a_s}
\newcommand{\muf}{\mu_F}
\newcommand{\mur}{\mu_R}
\newcommand{\mz}{m_z}

\newcommand{\eq}[1]{Eq.\ (\ref{#1})}
\newcommand{\fig}[1]{Fig.\ [\ref{#1}]}
\newcommand{\tab}[1]{Tab.\ [\ref{#1}]}

\newcommand{\pb}[1]{{\color{black}  #1}}
\newcommand{\mk}[1]{{\color{black}  #1}}

\title{{\boldmath 
Threshold resummation for $Z$-boson pair production at NNLO+NNLL}}

\author[a,b]{Pulak Banerjee\footnote{
Most of the work has been carried out at IIT Guwahati.
},}
\author[a]{Chinmoy Dey,}
\author[a]{M. C. Kumar }
\author[a]{and Vaibhav Pandey}
\affiliation[a]{Department of Physics, 
	Indian Institute of Technology Guwahati, Guwahati-781039, India}
\affiliation[b]{Istituto Nazionale di Fisica Nucleare, Gruppo collegato di Cosenza,
I-87036 Arcavacata di Rende, Cosenza, Italy }
\emailAdd{pulak.banerjee@lnf.infn.it}
\emailAdd{d.chinmoy@iitg.ac.in}
\emailAdd{mckumar@iitg.ac.in}
\emailAdd{vphiitg@iitg.ac.in }
\abstract{
The production of a pair of on-shell $Z$-bosons is an important process at the Large Hadron Collider(LHC). Owing
to its large production cross section at the LHC, this process is very useful for SM precision studies,
electroweak symmetry breaking sector as well as to unravel the possible new physics. In this work, we have performed the threshold resummation
of the large logarithms that arise in the partonic threshold limit $z \to 1$, up to Next-to-Next-to-Leading
Logarithmic (NNLL) accuracy. The presence of the two-loop contributions in the process dependent
resummation coefficient $g_0$ makes the numerical computation a non-trivial task. 
After matching the resummed predictions to the Next-to-Next-to-Leading order (NNLO) fixed order
results, we present the invariant mass distribution to NNLO+NNLL accuracy in QCD for the current
LHC energies. We find that in the high invariant mass region ($Q=1$ TeV), while the NNLO corrections
are as large as $83\%$  with respect to the leading order, the NNLL contribution enhances the cross 
section by additional few percent, about $4\%$ for $13.6$ TeV LHC.
In this invariant mass region, the conventional scale uncertainties in the fixed order results get
reduced from $3.4\%$ at NNLO to about $2.6\%$ at NNLO+NNLL, and this reduction is expected to be more for higher $Q$ 
values.

}

\begin{document}

\maketitle

\section{Introduction} \label{sec:introduction}
\input{introduction.tex}

\section{Theoretical Framework} \label{sec:theory}
\input{theory.tex}

\section{Numerical Results}\label{sec:numerics}
\input{discussion.tex}

\section{Summary}\label{sec:conclusion}
\input{conclusion.tex}

\section*{Acknowledgements}
 The research work of M.C.K. is supported by SERB Core 
Research Grant (CRG) under the project CRG/2021/005270. The
work of P.B. is supported by the INFN/QFT@COLLIDERS project (Italy).
The authors would like to thank A. H. Ajjath, M. Bonvini, L. Buonocore, G. Das and V. Ravindran for useful discussions. 
We acknowledge National Supercomputing
Mission (NSM) for providing computing resources of ‘{\tt PARAM Kamrupa}’ at IIT Guwahati, which is implemented by C-DAC and supported by the Ministry of Electronics and Information Technology (MeitY) and Department of Science and Technology (DST), Government of India,
where most of the computational work has been carried out. 

\pagebreak
\input{appendix.tex}

\bibliographystyle{JHEP}
 \bibliography{qqbZZ}
\end{document}

%% file: introduction.tex
The production of a pair of massive gauge bosons ($Z$) at the Large Hadron Collider (LHC) 
is an important process which has been studied very well, both theoretically and experimentally. This process
offers very clean signals that can be used to test the prediction of the Standard Model (SM) precisely, thanks
to their moderately large production cross sections at the current LHC energies. The process can also be
used for testing the SM electroweak symmetry breaking mechanism as well as in the study of fundamental weak 
interactions among elementary particles. Owing
to the large coupling between the Higgs and massive gauge bosons, the process plays an important role in the Higgs sector. One of the important decay modes of Higgs is to a pair of massive gauge bosons, either $ ZZ $ or $ WW $.

The experimental signature of this process typically involves either four charged leptons, two leptons plus 
missing energy or two leptons plus two jets or four jet events. Out of these, the decay to four charged leptons
provides a very clean signal in the collider experiments, which led to the measurement of these final states
both at the ATLAS and CMS experiments for different center of mass energies e.g. 5.02 TeV \cite{CMS:2021pqj}, 7 TeV \cite{ATLAS:2011ahl,ATLAS:2012bra,CMS:2012exm,CMS:2015qgb}, 8 TeV \cite{CMS:2015qgb,CMS:2013piy,CMS:2014xja,ATLAS:2015rsx,ATLAS:2016bxw,CMS:2018ccg}, 13 TeV \cite{CMS:2018ccg,CMS:2016ogx,ATLAS:2017bcd,CMS:2017dzg,ATLAS:2019xhj,CMS:2020gtj} and 13.6 TeV \cite{2024138764}.
Such measurements can be used to probe the trilinear gauge couplings as well as to probe the possible hidden
new physics . From the theoretical point of view, a similar study can easily be extended to the production
of a pair of new massive gauge bosons.

Owing to the importance of this process, a precise knowledge of this process, specifically the production 
cross sections as well as various kinematic distributions at the current LHC and future high energetic hadron
colliders, is very important. In the perturbative quantum chromodynamics (pQCD), 
the leading order (LO) predictions for this process have been
available for a long time \cite{PhysRevD.43.3626,Mele:1990bq,Zecher:1994kb,Ohnemus:1994ff}. It is well known that the LO predictions are unreliable and are contaminated
with large theoretical uncertainties. The next-to-leading order (NLO) calculations in pQCD were obtained for both
on shell as well as off shell $Z$-bosons decaying to a pair of leptons. The NLO QCD
corrections for this process can be found in Refs. \cite{Ohnemus:1992jn,Baur:1993ir,Baur:1997kz,Dixon:1998py,Dixon:1999di,Campbell:2011bn}. The NLO QCD results, matched with parton shower (NLO+PS) has been studied in Monte Carlo programs such as POWHEG \cite{Melia:2011tj,Nason:2013ydw} and aMC@NLO \cite{Frederix:2011ss}. The $ ZZ $ production at the LHC has been analyzed in the context of beyond Standard Model scenarios as well \cite{Agarwal:2009xr}.
It is to be noted that at the lowest order in the perturbation theory, this is a quark-antiquark-initiated process similar to the Drell-Yan (DY) production of dileptons.  However, the $Z$-boson pair production process has more similarities to the diphoton production process, 
as both involve identical particles in the final state, and both are $t$ and $u$ channel 
processes, whereas DY is an s-channel process ($s$,$t$, and $u$ being the Mandelstam variables). For the diphoton production process even at LO, simple
kinematic cuts are required to avoid divergences in the forward region, whereas for the case of $ZZ$ production, the mass of the $Z$-boson avoids
such divergences, and hence the total production cross section is finite even in
absence of any kinematic cuts. However, in the high invariant mass region, or in the region where $Z$-boson
carries much larger kinetic energy compared to the rest mass energy, both the processes can have similar behavior
in the cross sections, the special difference being the isolation algorithm to be used in the case of diphoton production process.

Precision studies entail going beyond NLO.
However, for the $Z$-boson pair production process, even the NNLO results
are challenging for both analytical as well numerical calculations. The full NNLO calculations have been
carried out in \cite{Binoth:2009wk,Campanario:2014ioa,Gehrmann:2015ora} in QCD for the quark annihilation process.  It is also worth noting that the NNLO 
corrections are computed for both the on shell Z-boson \cite{Cascioli:2014yka,Heinrich:2017bvg} case as well as for off shell $Z$-bosons \cite{Grazzini:2015hta,Kallweit:2018nyv} followed by their 
decay to lepton final states. With the availability of two loop helicity amplitudes \cite{Gehrmann:2015ora}, the differential 
distributions for the latter case also became possible. Fiducial cross sections and distributions are also available
 for the vector boson pair production processes \cite{Grazzini:2015hta,Grazzini:2017ckn}.
Additionally, the LO process in the gluon fusion channel, $gg \to ZZ$, also contributes at ${\cal O}(\alpha_s^2)$.  In the low invariant mass region, the gluon fluxes for LHC energies are very large 
and hence the contribution from this channel near the hadronic threshold region is very crucial
and cannot be neglected. Going beyond NNLO for $Z$-boson pair production processes is a challenging task. On the other hand, for DY and Higgs production this has been achieved. Recently the full
N$^3$LO results for DY have become available and can be found in Refs. \cite{Duhr:2020seh,Duhr:2020sdp,Duhr:2021vwj,Baglio:2022wzu} and soft-virtual (SV) \cite{AH:2020cok,Das:2022zie,Das:2024auk} and  next-to-soft-virtual (NSV) \cite{AH:2020iki,AH:2021kvg,AH:2022lpp} threshold resummations of DY-type processes are available.
The Higgs production through gluon fusion channel N$^3$LO results are available in \cite{Baikov:2009bg,Gehrmann:2010tu,Anastasiou:2013srw,Anastasiou:2015vya,Dulat:2017prg} and SV \cite{Bonvini:2014joa,Bonvini:2016frm,Ahmed:2015qda,Ahmed:2016otz} and NSV \cite{AH:2020iki,AH:2021vdc,Bhattacharya:2021hae} threshold resummation results are also available. For Higgs production via bottom quark annihilation,  SV \cite{AH:2019phz} and NSV \cite{Das:2024pac} resummation results are also available.
Rapidity resummation for DY processes are available for SV \cite{Banerjee:2018vvb,Das:2023bfi} and NSV \cite{AH:2020qoa,Ahmed:2020amh,AH:2021vhf,Ravindran:2023qae}.
Rapidity distribution for the Higgs production at the threshold to third order in the gluon fusion channel can be found in \cite{Ahmed:2014uya}. 
For the DY type processes, the threshold results up to third order can be found in \cite{Ahmed:2014cla, Kumar:2014uwa, Das:2022zie}. 
This also enables the evaluation of the resummed parton distribution functions (PDFs), which are important for high precision studies in the large and small $x$ regions. To see the impact of resummed PDF's on DY and Higgs production; see Refs. \cite{Bonvini:2015ira,Rottoli:2017ifw}. Rapidity resummation for gluon fusion channel Higgs production is available for SV \cite{Banerjee:2017cfc} and NSV \cite{AH:2020qoa,Ravindran:2022aqr} cases. Rapidity resummation for Higgs production via bottom quark annihilation is also available at NNLO+NNLL in Ref. \cite{Das:2023rif}. 
Nevertheless, for the $Z$-boson pair in the final state, there has been a tremendous effort to go beyond NNLO. Transverse momentum resummation for vector boson pair production is available up to NNLO+N$^3$LL \cite{Grazzini:2015wpa,Campbell:2022uzw}. 
The parton shower matched with NNLO (NNLO+PS) are recently studied using the $ \rm MINNLO_{PS} $ \cite{Buonocore:2021fnj} method for the $ZZ$ production.
The LO matching with parton shower results are available for the gluon fusion channel as well in Ref. \cite{Binoth:2008pr}.
The NLO corrections to this channel have also become available \cite{Caola:2015psa,Grazzini:2018owa,Grazzini:2021iae,Agarwal:2024pod}. However, their
contribution in the high invariant mass region ($\ge 2000$ GeV) becomes much smaller than those in the
quark annihilation channel.
The NLO results matching with parton shower (NLO+PS) results for the gluon fusion channel with massless quarks are also available \cite{Alioli:2016xab}. Finally, at this precision level, the electroweak corrections cannot be 
ignored for precision studies and the NLO EW corrections to this process have been computed in Refs.  \cite{Bierweiler:2013dja,Grazzini:2019jkl,Denner:2021csi}. Using SCET formalism, threshold resummation for vector boson pair
production is available up to NLO+NNLL \cite{Wang:2014mqt}. 

While the threshold resummation for the final state onshell $Z$-bosons
has been achieved to NLO+NNLL level, it is necessary to go beyond this accuracy.
The $Z$-boson pair production cross section has been measured at the LHC experiments using $13$ TeV data obtained from the $137$ fb$^{-1}$ luminosity. 
The measured total production cross section is $17.4$ pb with an error of about $0.8$ pb \cite{CMS:2020gtj}. 
The invariant mass distribution also has been measured up to 1 TeV region. 
With the upcoming High-Luminosity LHC (HL-LHC), the integrated luminosity is expected to reach $3000-4000$ fb$^{-1}$, which allows the measurement of $ZZ$ events in this TeV region with more statistical data. 
Considering  DY production of a lepton pair, the threshold resummation is known till N$^3$LO + N$^3$LL  accuracy.
In Ref. \cite{Das:2022zie}, invariant mass regions around 3 TeV were studied, which is also the same mass range for which experimental results exist in Ref. \cite{CMS:2018mdl}.
The experimental uncertainty reported for the DY process for the invariant mass in the Q range $1000-1500$ GeV at $7$ TeV LHC \cite{ATLAS:2013xny}, is around $50 \%$, which decreases by $25\%$ for the $13$ TeV in the same Q-range \cite{CMS:2018mdl}.
From the theory side, for the same $ Q=1.5$ TeV, the scale uncertainties decrease from $0.77\%$ at NNLO level to about $0.28\%$ at NNLO+NNLL level after incorporating threshold resummation~\cite{Das:2022zie}.

For the planned FCC-hh \cite{FCC:2018vvp}, the increase in parton fluxes along with the integrated luminosity of $20-30$ ab$^{-1}$ can enhance the invariant mass distributions by a few orders of magnitude, thus allowing the measurement of the $ZZ$ events beyond $1$ TeV region even more precisely. 
At this level of experimental precision, adequate theoretical predictions are necessary. 
In this work, we attempt to increase the theoretical precision by performing the resummation at NNLL accuracy and give the results after matching them to the available fixed-order results at the NNLO level.

Our paper is organized as follows: we present the theoretical framework in Sec. \ref{sec:theory}. Details of the
phenomenological analysis and the numerical results are presented in Sec. \ref{sec:numerics}. Finally,
in Sec. \ref{sec:conclusion}, we summarize our observations.

%% file: theory.tex
The hadronic cross section for $Z$-boson pair production  
can be written in terms of its partonic counterpart as follows:
\begin{align}\label{eq:had-xsect}
 \frac{d\,\sigma}{d\, Q}  =
\sum_{a,b= \{q, \bar{q}, g\}}\int_0^1 dx_1\int_0^1 dx_2 \,\,f_{a}(x_1,\mu_F^2)\,
f_{b}(x_2,\mu_F^2)
\int_0^1 dz~ \delta \left(\tau-z x_1 x_2 \right)
 \frac{d\,\hat\sigma_{ab}}{d\, Q} \,.
\end{align}
Here $Q$ is the invariant mass of the final states. The hadronic and partonic threshold variables $\tau$
and $z$ are defined as
\begin{align}
\tau=\frac{Q^2}{S}, \qquad z= \frac{Q^2}{s} \,,
\end{align}
where $S$ and $s$ are the hadronic and partonic
center of mass energies, respectively.
$\tau$ and $z$ are thus related by $\tau = x_1 x_2 z$.

The leading order DY type parton level process has the generic form
\begin{align}
	q(p_1) + \bar{q}(p_2) \to Z(p_3) + Z(p_4).
	\label{eq:parton}
\end{align}

The LO cross-section $\hat{\sigma}^{(0)}_{q \bar{q}}$
for $Z$-boson pair production can be written as, 
\begin{align}
	\frac{d\hat{\sigma}^{(0)}_{q \bar{q}}}{dQ} = \frac{1}{2 s}\int dPS_2 ~\mathcal{M}_{(0,0)},
	\label{zzbornVP}
\end{align}
where $dPS_2$ is the two-body phase space integration and  $ \mathcal{M}_{(0,0)} $ is the born amplitude, which in $ d=4-2\epsilon $ dimensions is given below,
\begin{align}\label{eq:born}

		\mathcal{M}_{(0,0)} = & B_{f} \,  {\rm N} (c_a^4 + 6 c_a^2 \, c_v^2 + c_v^4)
			  \frac{4}{t^2 \, u^2}					  
			  \Biggl\{
			  - m_{z}^4 t^2
			  + 8 m_{z}^4 t u 
			  + t^3 u 
			  - m_{z}^4 u^2
			  + t u^3
			  - 4 m_{z}^2 t u(t + u)
			  \nn
			  & + \epsilon \, (
			  - 2 m_z^4 t^2
			  - 6 m_z^4 t u
			  - 2 t^3 u
			  + 2 m_z^4 u^2
			  + 2 t^2 u^2
			  - 2 t u^3
			  + 4 m_z^2 t u ( t + u)
			   )
			   \nn
			 &  + \epsilon^2 \, (
			  - m_z^4 t^2
			  - 2 m_z^4 t u 
			  + t^3 u
			  - m_z^4 u^2
			  + 2 t^2 u^2
			  + t u^3
			  )
			  \Biggr\}.
\end{align}

In \eq{eq:born} the kinematical variables $ s,t, $ and $ u $ are defined as
\begin{align}
	(p_1+p_2)^2 = & s\,,~~~
	(p_1-p_3)^2 =  t\,,~~~ 
	(p_2-p_3)^2 =  u~~~
{\rm and} ~~~
	s + t + u = 2 \mz^2,
\end{align}
\begin{align*}
	B_{f} = \frac{(4 \pi \alpha)^2}{4 {\rm N}^2} \,,~~~
	c_{v} = \frac{\big( T_3^f - 2~ {\rm sin}^2\theta_{\rm w} Q_{f} \big)}{2~ {\rm sin}\theta_{\rm w} {\rm cos}\theta_{\rm w} } \,,~~~
	c_{a} = \frac{1}{4~ {\rm sin}\theta_{\rm w} {\rm cos}\theta_{\rm w} }\,,
\end{align*}
where the $Q_f$ and $T_3^f$ are the electric charge and the third component of weak isospin of the fermion $f$ and $\theta_{\rm w}$ is the weak mixing angle. Here, $\mz$ is the mass of the $Z$-boson, N is the SU(N) color, and $\alpha$ is the fine structure constant. Here, we have dealt with $\gamma_5$ in $d$-dimension using Larin's prescription \cite{Larin:1993tq}. \\
Beyond LO, the partonic cross-section receives corrections originating from virtual and real contributions. It is interesting to study the cross section in the soft limit, which is defined by, $z \rightarrow 1$. This means that the initial partonic center of mass energy is almost used to produce the final state pair of $Z$-bosons, and small energy is left to produce soft partons.  In this limit, the partonic cross section can be organized as follows:
\begin{align}
 \frac{d\,\hat\sigma_{ab}}{d\, Q} =
	\frac{d\,\hat\sigma^{(0)}_{ab}}{d\, Q}\Big(
\Delta_{ab}^{\rm sv}\left(z,\muf^2\right)
+ \Delta_{ab}^{\rm reg}\left(z,\muf^2\right)
\Big) \, \,.
\label{eq:bornG}
\end{align}

The term
$\Delta_{ab}^{\rm sv}$ is known as the soft-virtual (SV)
partonic coefficient and captures all the singular terms
in the $z \to 1$ limit. Only quark-antiquark or gluon-gluon subprocesses contribute to this SV cross section. The
$\Delta_{ab}^{\rm reg}$ term contains regular (hard) contributions
in the variable $z$. Both these contributions are expanded in a perturbative series of the strong coupling constant. In our work, we consider such an expansion up to NNLO in QCD.
It is to be noted that the overall
normalization factor
$d\hat{\sigma}^{(0)}_{ab}/dQ$ depends on the process
under study.

The singular part of the partonic coefficient
has a universal structure which gets
contributions from the underlying hard form
factor
\cite{Moch:2005tm,
Moch:2005id,
Baikov:2009bg,
Gehrmann:2010ue,
Gehrmann:2014vha},
mass factorization kernels
\cite{Moch:2004pa,Vogt:2004mw,Blumlein:2021enk}
and soft radiations
~\cite{ Sudakov:1954sw, Mueller:1979ih, Collins:1980ih, Sen:1981sd, Sterman:1986aj, Catani:1989ne, Catani:1990rp,Kidonakis:1997gm,Kidonakis:2003tx,Ravindran:2005vv, Ravindran:2006cg,Moch:2005ba,Laenen:2005uz,Kidonakis:2005kz,Idilbi:2006dg}.
According to the KLN theorem, these infrared divergences, when regularized and combined, yield finite contributions.
After the infrared cancellation, the finite part of these
has the universal structure in terms of
$\delta(1-z)$ and plus distributions
${\cal D}_i = [\ln(1-z)^i/(1-z)]_+$.
In the threshold limit, $z \to 1$, the plus distributions contribute dominantly to the SV cross section.
These large distributions can be resummed to all
orders in the threshold limit.
Threshold resummation is conveniently performed
in the Mellin ($N$) space where the convolution
structures become simple product.

The partonic coefficient in the Mellin space is
organized as follows:
\begin{align}\label{eq:resum-partonic}
	 \hat{\sigma}_N^{\rm N^{n}LL}
        = \int_0^1 \df z ~ z^{N-1} \Delta^{\rm sv}_{a b}(z)
        \equiv g_{0} \exp \left( G_N \right).
\end{align}
The factor $g_{0}$ is independent of the Mellin variable,
whereas the threshold enhanced large logarithms ($\ln^{i} N$ in Mellin space)
are resummed through the exponent $G_{N}$.
The resummed accuracy is determined through the
successive terms from the exponent $G_N$ which up to
NNLL takes the form,
\begin{align}\label{eq:gn}
        G_N =
        \ln (\Nbar) ~\gbar_1(\Nbar)
        +\gbar_2(\Nbar)
        +\as ~\gbar_3(\Nbar)+ ...
\end{align}
where $\Nbar = N \exp (\gamma_E)$.
These coefficients are universal
and only depend on the partonic flavors being
either quark or gluon. Their explicit form can be
found in Refs. \cite{Catani:2003zt,Moch:2005ba}.
In order to achieve complete resummed accuracy
one also needs to know the $N$--independent
coefficient $g_0$ up to sufficient accuracies.
In particular, up to NNLL, it takes the form,
\begin{align}\label{eq:g0}
        g_0
        =
        1
        + \as ~g_{_{01}}
        + \as^2 ~g_{_{02}}+...
\end{align}
where $\as=\frac{\alpha_s}{4 \pi}$ and $\alpha_s$ is the strong coupling constant. Using the universal $G_N$  and the process dependent $g_0$, resummation for two Higgs boson production in the gluon fusion channel at N$^3$LO + N$^3$LL has been achieved in \cite{AH:2022elh}. 

It is also possible to resum part (or all)
of the $g_0$ by including them in the exponent
\cite{Bonvini:2014joa,Bonvini:2016frm,Eynck:2003fn,Das:2019btv,Ajjath:2020rci},
which, however, have a subleading effect as these
contributions are not dominated in the threshold
region.

The $N$-independent coefficient $ g_{0} $ is computed based on the formalism given in Ref. \cite{Ahmed:2020nci}, the expression for $ g_{01} $ and $ g_{02} $ are given in Appendix \ref{appendixa}. The $ g_{01} $ requires one-loop virtual computations ($ \mathcal{M}_{(0,1)} $); we have computed the amplitude $ \mathcal{M}_{(0,1)} $ using our inhouse \texttt{FORM} \cite{Ruijl:2017dtg} code and the expression is given in Appendix \ref{appendixa}.
%
To obtain the results in $z$ space, one needs to do the Mellin inversion as,
\begin{align}
	\frac{d\sigma^{\rm N^{n}LL}}{d Q}
	=&
	\frac{d\hat{\sigma}^{(0)}}{d Q}
	\sum_{a,b \in \{q,\bar{q}\}}
	\int_{c-i\infty}^{c+i\infty}
	\frac{\df N}{2\pi i}
	\tau^{-N}
	f_{a,N}(\muf)
	f_{b,N}(\muf)
		~ \hat{\sigma}_N^{\rm N^{n}LL} \,. 
\end{align}

This complex integral contains the Landau pole at
$N =\exp\big(1/(2 \as \beta_0) -\gamma_E\big)$, which makes the choice of contour very important.
The Mellin inversion is performed \cite{Vogt:2004ns} along the
contour $N=c + x~ \exp(i\phi)$, where $x$ is real variable. Following the \textit{minimal prescription} \cite{Catani:1996yz}, we choose
the value of $c$ such that all the singularities except the Landau pole lie on the left side of the integration contour.
For numerical results we choose $c=1.9$ and $\phi=3\pi/4$. 
\mk{It may be noted that the results are dependent on the choice of the prescription. Several studies have been done in the past in this direction \cite{Berger:1997gz,Kidonakis:2000ui,Forte:2006mi,Kidonakis:2018ybz,Hinderer:2018nkb}.
It is to be noted that the there are other prescriptions to deal with the Landau pole like Borel presciption \cite{Bonvini:2016frm}. Although these prescriptions differ by subleading terms, they give similar numerical results in the high $Q$ region \cite{Bonvini:2016frm}. 
}

Finally, the matched results can be written as,
\begin{align}\label{eq:matched}
	\frac{d \sigma^{\rm N^{n}LO+N^{n}LL}}{d\, Q}
        =&
	\frac{d\sigma^{\rm N^{n}LO}}{d\, Q}
        +
	\frac{d\hat{\sigma}^{(0)}}{d\, Q}
        \sum_{a,b \in \{q,\bar{q}\}}
        \int_{c-i\infty}^{c+i\infty}
        \frac{\df N}{2\pi i}
        \tau^{-N}
        f_{a,N}(\muf)
        f_{b,N}(\muf)
        \nn
        &\times
        \bigg(
                \hat{\sigma}_N^{\rm N^{n}LL}
                -
                \hat{\sigma}_N^{\rm N^{n}LL} \bigg|_{\rm tr}
        \bigg).
\end{align}

In the above equation, $f_{a,N}$ in the Mellin transformed
PDF, which one can obtain using publicly available code such as QCD-PEGASUS \cite{Vogt:2004ns}. However, for numerical applications, it can also be approximated by employing the $z$-space PDF following Refs. \cite{Catani:2003zt,Catani:1989ne}.
The last term in the bracket of \eq{eq:matched} indicates the truncation of the resummed partonic coefficient \eq{eq:resum-partonic}, which avoids double counting the regular terms already present in the fixed order.

Before we proceed to the numerical results, we add a few comments on the extraction of the resummation coefficient, $ g_{02} $, which is a process-dependent quantity.
To obtain the full $ g_{02} $, we follow the procedure given in Ref. \cite{Ahmed:2020nci} and make use of the two-loop ($ \mathcal{M}_{(0,2)} $) amplitudes as well as the one-loop squared amplitudes, ($\mathcal{M}_{(1,1)}$). The two-loop virtual amplitudes $ \mathcal{M}_{(0,2)} $ are reconstructed using the {\tt VVamp} package \cite{Gehrmann:2015ora},  while the $ \mathcal{M}_{(1,1)} $ is obtained by squaring
$ \mathcal{M}_{(0,1)} $.
After performing the UV renormalization, the IR pole structure of these amplitudes is then verified using Catani's I-operator formalism \cite{Catani_1998}.
The finite parts of the two-loop amplitudes are quite large in size. These amplitudes are then simplified and optimized using our inhouse-developed \texttt{FORM} routines. We have used {\tt handyG} \cite{Naterop:2019xaf} for numerical evaluations of the generalized polylogarithms in these one-loop and two-loop amplitudes. Finally, these two-loop amplitudes have been cross-checked numerically for different phase space points with those implemented in the \texttt{MATRIX} \cite{Grazzini:2017mhc,Gehrmann:2015ora,Denner:2016kdg,Cascioli:2011va,Buccioni:2019sur,Buccioni:2017yxi}. The resummation formalism used here is developed in \cite{Catani:1989ne,Banerjee:2017cfc,Banerjee:2018vvb} and has been successfully used in the past in calculating NNLO$+$NNLL and N$^{3}$LO+N$^{3}$LL  for Drell-Yan processes \cite{Das:2020pzo,Das:2022zie, AH:2020cok, Ahmed:2015qda}. All the anomalous dimensions required for computing the resummed results are given in the Appendix \ref{appendixa}.

%% file: discussion.tex
In this section, we present the numerical results for the $Z$-boson pair production process at the LHC.
For the numerical computation, we take the fine structure constant to be $\alpha = 1/132.233193$.
The mass of the weak gauge bosons $\mz = 91.1876$ GeV, $ m_{w}=80.385 $ GeV. 
The Weinberg angle is $\text{sin}^2\theta_\text{w} = (1 - m_w^2/\mz^2) = 0.222897223$.
This corresponds to the weak coupling $G_F = 1.166379\times 10^{-5} \text{ GeV}^{-2}$. 
The default choice of centre mass energy of the incoming protons is $13.6$ TeV.
Unless specified otherwise, in our numerical analysis, we use MSHT20 \cite{Bailey:2020ooq} parton
distribution functions (PDFs) throughout, taken from the {\tt LHAPDF} \cite{Buckley:2014ana}.
The LO, NLO and NNLO cross sections are obtained by convoluting the respective coefficient functions 
with MSHT20lo\_as130, MSHT20nlo\_as120 and MSHT20nnlo\_as118 PDF sets, using the central set (iset=0)
as the default choice.
The strong coupling constant $\as$ is taken from {\tt LHAPDF} \cite{Buckley:2014ana}, and it
varies order by order in the perturbation theory. For our analysis, we consider the number of light quark flavors as $\nf = 5$. For the fixed-order calculations, we have used
the package {\tt MATRIX} \cite{Grazzini:2017mhc}. The fixed-order results from MATRIX are checked against those obtained from the MadGraph package \cite{Alwall_2011} up to NLO and are found to be in good agreement.  The fixed-order NNLO results have been checked with the numbers quoted in the literature \cite{Cascioli:2014yka, Grazzini:2015hta}.
The resummation results are obtained using the inhouse developed numerical code. 
The unphysical renormalization and factorization scales are set to $\mu_R = \mu_F = Q$, where $Q$ is the invariant mass of the $Z$-boson pair production in the final state. 
The scale uncertainties are estimated by varying the unphysical scales in the range so that $|\,\text{ln}(\mu_R/\mu_F)\,| \le \text{ln}\,2$. The symmetric scale uncertainty is calculated from the maximum of the absolute deviation of the cross section from that
obtained with the central/default scale choice. 

\pb{We first comment on how much the SV results are in comparison to the full fixed-order cross section. 
The SV results contain distributions, which contribute significantly in the threshold region. 
We find that at the SV corrections at the one-loop level are about about 79\% of the full first order correction at Q=595 GeV, and this contribution increases to 94\% at Q= 995 GeV. 
Similarly, at NNLO level, the second-order SV contributions contribute about $57\%$ of the full second-order result at Q=595 GeV, and this contribution increased to $80\%$ at $Q= 995$ GeV.}
\mk{A similar observation of large SV contributions has been reported in \cite{Ahmed:2014uya} in the context of Drell-Yan production of di-leptons.}
\mk{With the available ingredients i.e. (without the three loop virtual corrections) it is also possible to estimate the size of the soft-gluon contributions at three loop level. 
The SV corrections thus computed at three loop level are found to be around $0.2\%$ of LO.}

To estimate the impact of the higher-order corrections from FO and resummation, we define the following ratios
of the cross sections which are useful in the experimental analysis:

\begin{align}
K_{\text{ij}}
=
	\frac{\sigma_{\text{N}^i\text{LO}}}{\sigma_{\text{N}^j\text{LO}}}
\,,~
R_{\text{ij}}
=
\frac{\sigma_{\text{N}^i\text{LO} + \text{N}^i\text{LL}}}{\sigma_{\text{N}^j\text{LO}}}~
\text{ and }
~
L_{\text{ij}}
=
\frac{\sigma_{\text{N}^i\text{LO} + \text{N}^i\text{LL}}}{\sigma_{\text{N}^j\text{LO}+\text{N}^j\text{LL}}}
~
\text{ with } i, \, j= 0, 1 \text{ and } 2 \, \cdot
        \label{eq:ratio}
\end{align}
%



\begin{figure}[ht!]
	\centerline{
		\includegraphics[scale =0.8]{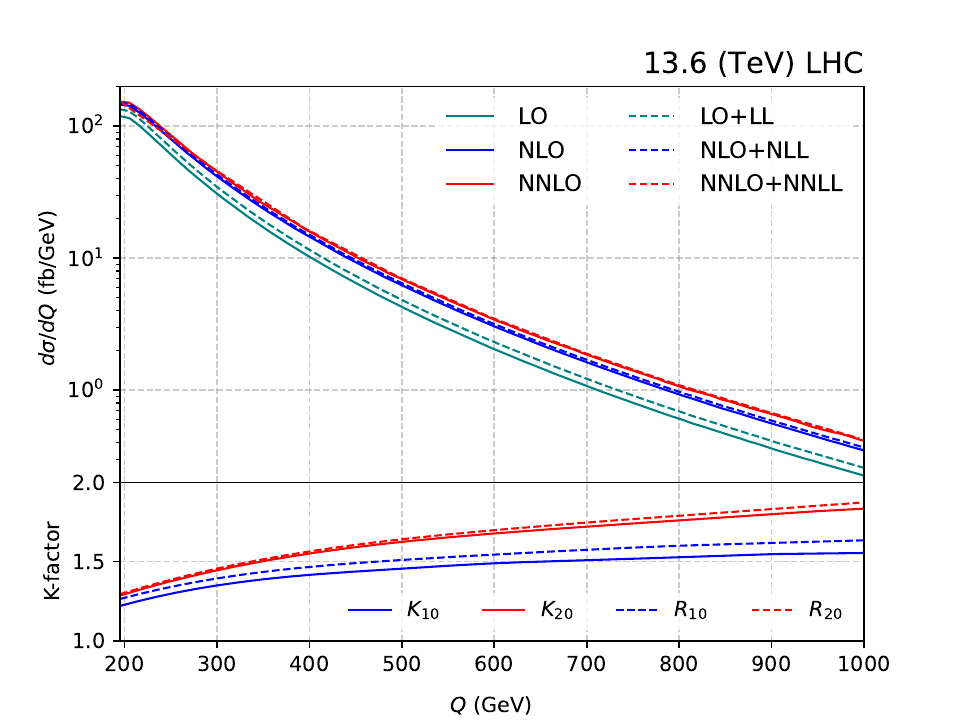}
	}
	\vspace{-2mm}
	\caption{\small{Resummed predictions for the $Z$-boson pair invariant mass distribution (upper panel) and the corresponding K-factors(lower panel) up to NNLO+NNLL.}}
	\label{fig:match_ZZ_inv}
\end{figure}

\begin{figure}[ht!]
%

        \centering
        \includegraphics[scale=0.5]{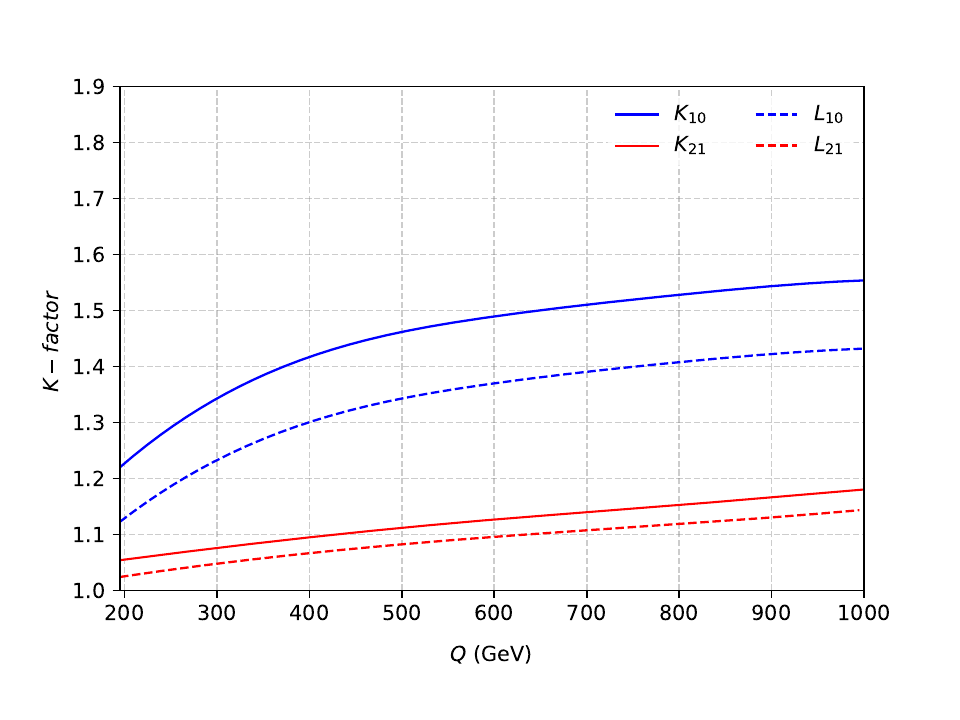}

    \vspace{-2mm}
    \caption{\small{K-factors for the fixed-order and resummed results for $Z$-boson pair production(see \eq{eq:ratio} for details).}}
    \label{fig:match_ZZ_kfac}
\end{figure}

\begin{figure}[ht!]
	\centerline{
		\includegraphics[scale =0.5]{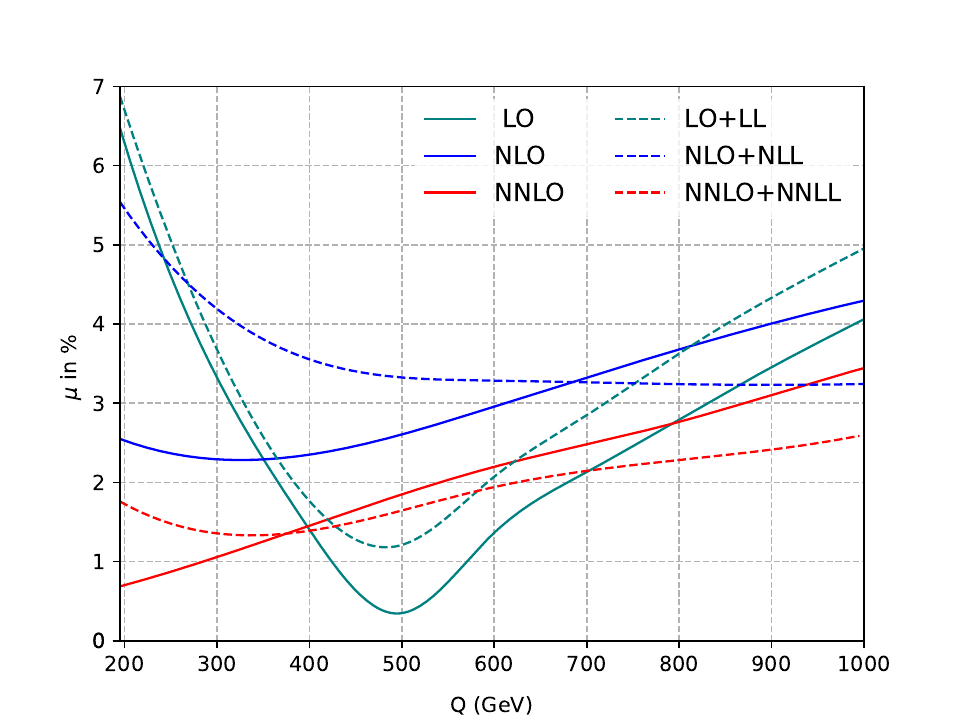}
	}
	\vspace{-2mm}
	\caption{\small{Seven point scale uncertainties for $Z$-boson pair production up to NNLO+NNLL.}}
	\label{fig:match_ZZ_mu}
\end{figure}

\begin{figure}[ht!]
	\centerline{
		\includegraphics[scale =0.5]{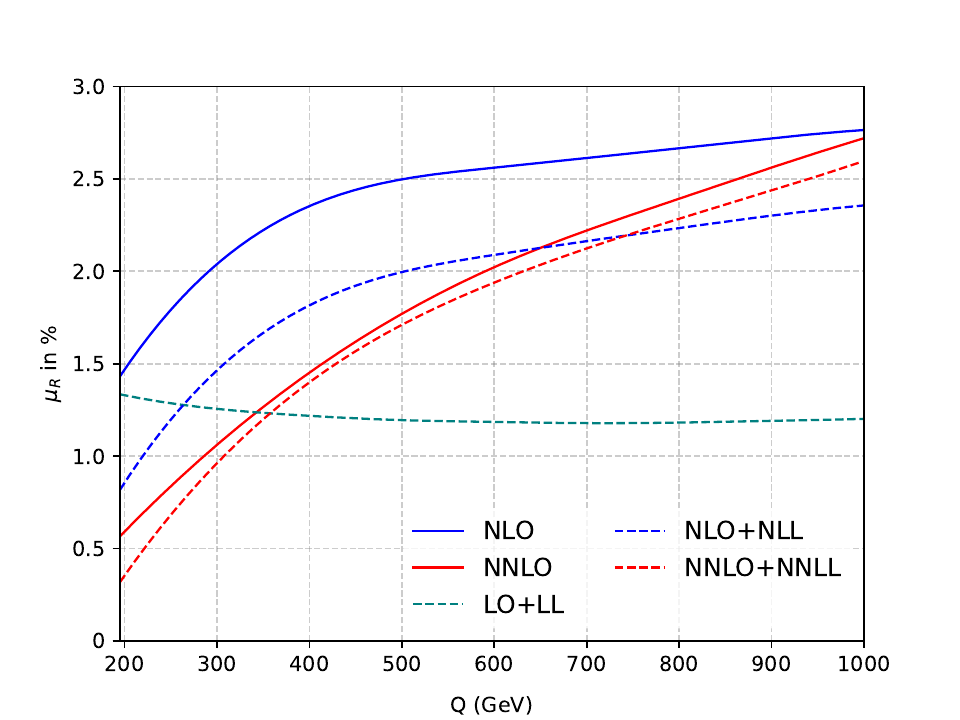}
		\includegraphics[scale =0.5]{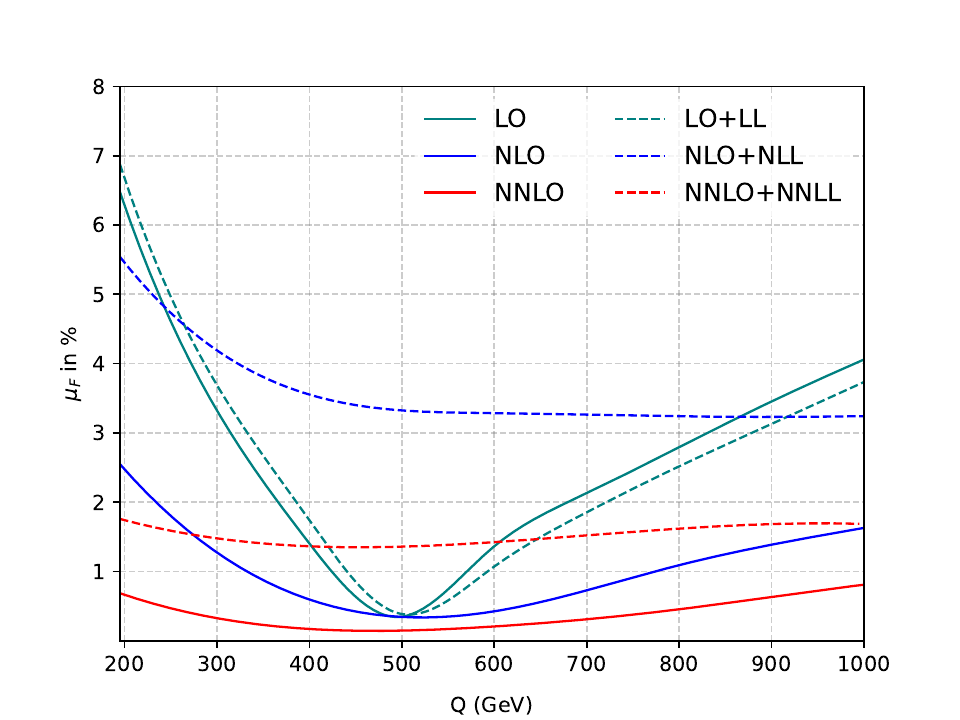}
	}
	\vspace{-2mm}
	\caption{\small{Renormalization(left) and factorization(right) scale uncertainties for $Z$-boson pair production up to NNLO+NNLL.}}
	\label{fig:match_ZZ_mur}
\end{figure}

\begin{figure}[ht!]
   \centering
   \begin{minipage}{0.45\textwidth}
        \centering
        \includegraphics[scale=0.5]{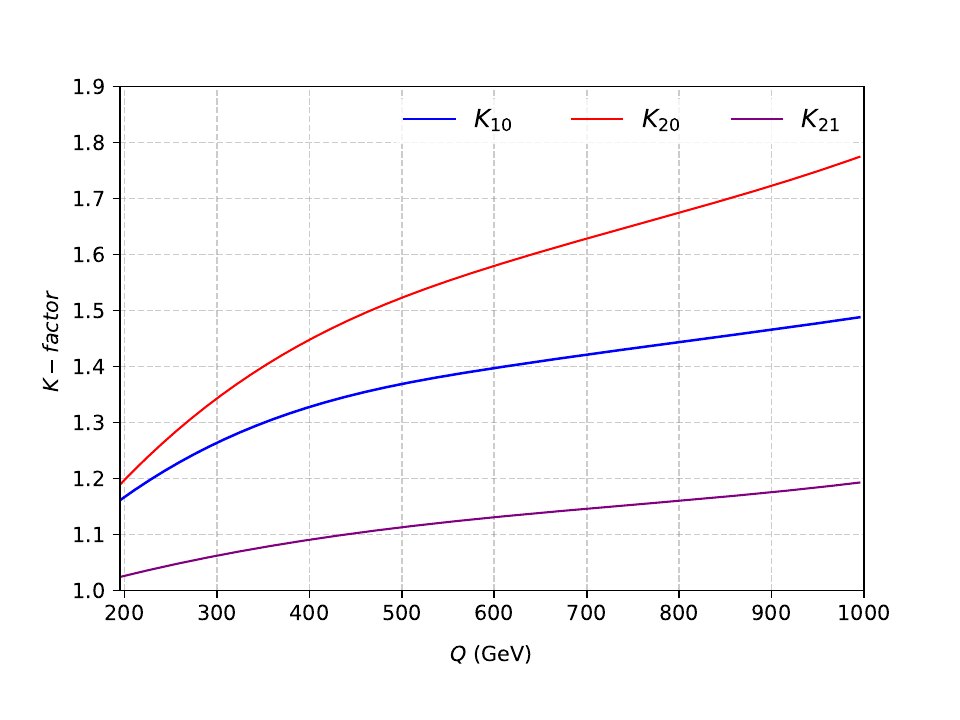}
    \end{minipage}
    \hfill
    \begin{minipage}{0.45\textwidth}
        \centering
        \includegraphics[scale=0.5]{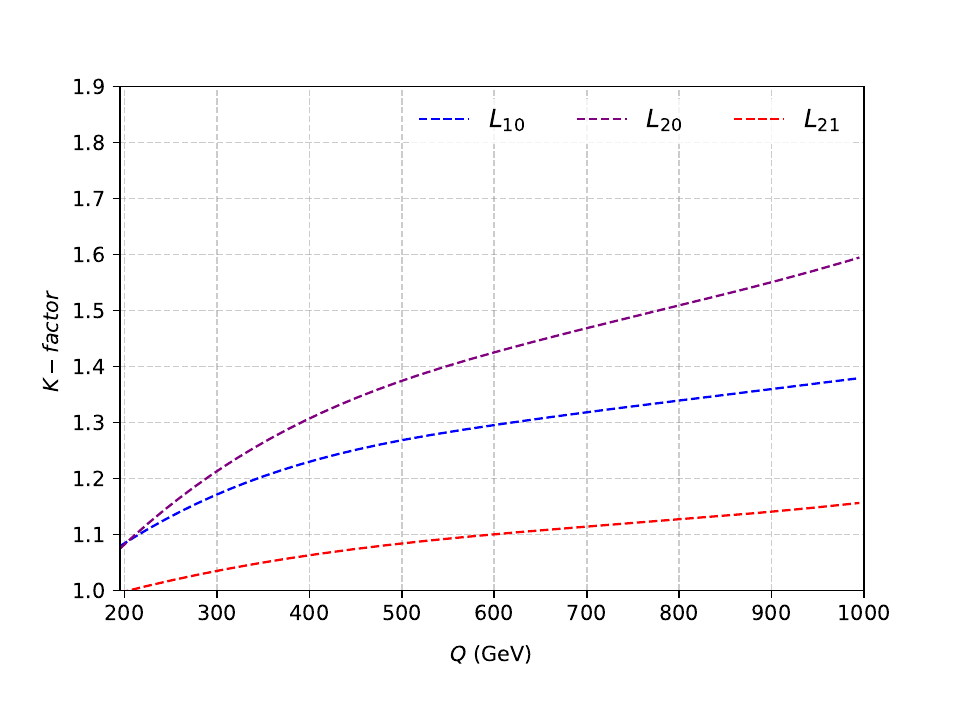}
    \end{minipage}
    
    \begin{minipage}{0.45\textwidth}
        \centering
        \includegraphics[scale=0.5]{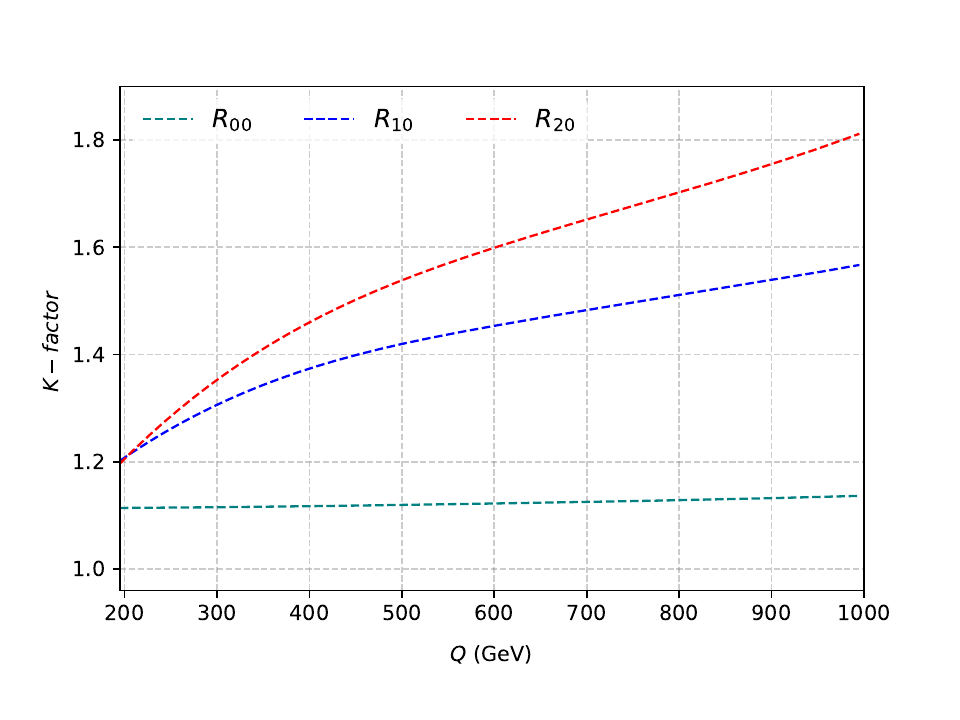}
    \end{minipage}
    \hfill
    \begin{minipage}{0.45\textwidth}
        \centering
        \includegraphics[scale=0.5]{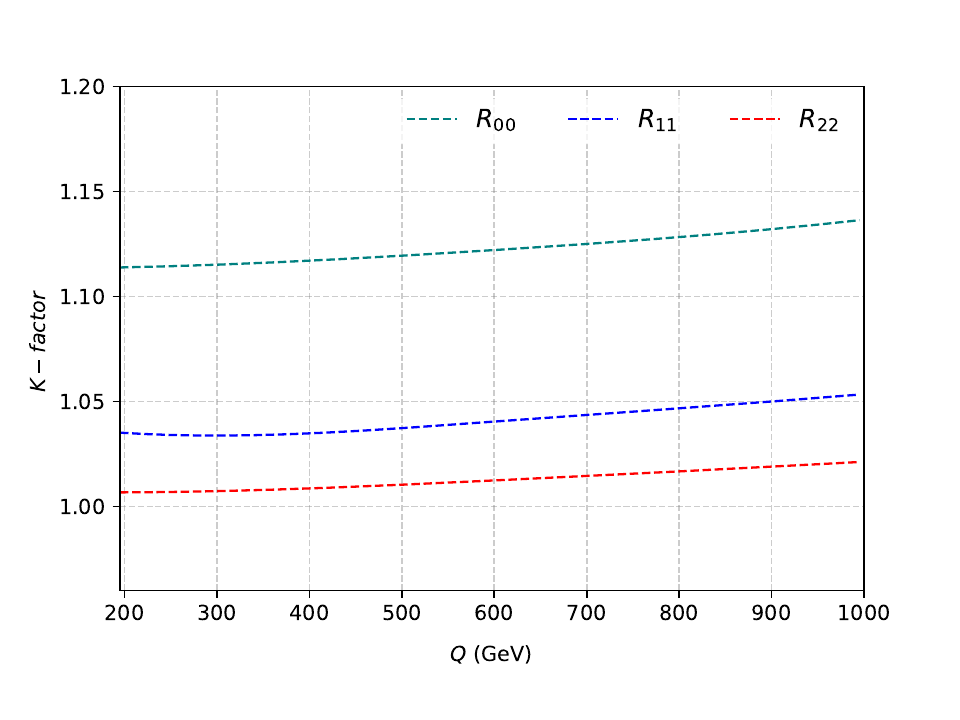}
    \end{minipage}

       \vspace{-2mm}
	\caption{\small{K-factors as defined in \eq{eq:ratio} but using same NNLO PDFs at various orders in the perturbation theory.}}
\label{fig:match_ZZ_kfacnnlo}
\end{figure}
In \fig{fig:match_ZZ_inv} , we present the fixed-order results for the invariant mass distribution of the $Z$-boson pair production 
from $2\mz$ to $1300$ GeV up to NNLO+NNLL in QCD. For the range of $Q$-variation considered here, the distribution
varies over three orders of magnitude. In the lower panel, the corresponding fixed-order as well as the
resummed K-factors, as defined in \eq{eq:ratio} are given. The NLO K-factor $K_{10}$ here varies from 
about $1.22$ at $Q=200$ GeV to about $1.58$ at $Q=1300$ GeV. As can be seen from \fig{fig:match_ZZ_inv}, the variation
of $K_{10}$ above $Q=500$ GeV is mild. However, the NNLO K-factor $K_{20}$ slowly but continuously increases
from about $1.29$ to about $1.94$  for the $Q$ range considered here. This clearly shows additional contributions 
coming from second-order corrections in the higher $Q$-region,
which are due to the real correction subprocesses like $qq^\prime \to ZZ qq^\prime$. 
A more detailed discussion of this kind of contributions can be found in the
Ref. \cite{Grazzini:2019jkl}.
The resummation has been achieved in the Mellin space where large logarithms of kind ln($\bar{N}$) have been resummed to NNLL accuracy as outlined in the text. From $R_{00}$, we see that the LL resummation enhances the LO results by about $15\%$ for $Q=1300$ GeV. 
While the NLO corrections are as big as  $58\%$, the corresponding
NLO+NLL results are about $68\%$ of LO for the same $Q$ region. 
The corresponding contribution from NNLO is about $94\%$
where the NNLL resummation adds an additional few percent, making the NNLO+NNLL contributions sizable, about $99\%$ of LO. 

To better estimate the size of higher-order corrections beyond a certain fixed-order as well as that of resummed contributions at 
different logarithmic accuracy, we present various K-factors of the kind $K_{i,i-1}$, $R_{i,i-1}$ and $L_{i,i-1}$, in \fig{fig:match_ZZ_kfac}.
We notice that, in general, $L_{ij}$ are smaller compared to the respective $K_{ij}$, indicating that the threshold logarithms of a given accuracy (LL, NLL, ...) capture a substantial contribution of the higher-order corrections.
Moreover, we notice that the gap between $K_{21}$ and $L_{21}$ is smaller than that between $K_{10}$ and $L_{10}$, demonstrating a nice convergence of the higher-order QCD corrections. Further, it is evident that the former gap 
is almost independent of $Q$ while the latter gap increases with $Q$. 
We also notice that while the $L_{10}$ is about as large as $1.45$, 
the $L_{21}$ is smaller and is about $1.18$. 
This indicates that the contribution of the second-order terms and the tower of further 
sub-leading logarithms (NNLL) that are not included in NLO+NLL are about $18\%$ of NLO+NLL, and are still non-negligible for precision studies.

The underlying theory uncertainties in these distributions up to NNLO+NNLL due to the variation of arbitrary factorization and renormalization scales are presented in \fig{fig:match_ZZ_mu}. Here, the complete
$7$-point scale variations, as discussed in the text, have been presented where the maximum uncertainty in the low Q-region at LO is about $6.5\%$. However, the inclusion of higher-order NLO and NNLO corrections reduce this uncertainty to as low as $2.6\%$ and to less than $1.0\%$ respectively, for $Q=200$ GeV. The general observation is that these scale uncertainties are found to increase with $Q$, which for the NNLO case are found to change from about $0.7\%$ to about $4.5\%$. 


For higher $Q$ values, the threshold logarithms are dominant and as a result of
resummation, the corresponding scale uncertainties are expected to be smaller than those
in the fixed-order ones. We see that the scale uncertainties increase from LO to LO+LL, this is because for the process under study there is
no $\mu_R$ scale dependence at LO. On the other hand, the $7$-point scale uncertainties 
at NLO+NLL (NNLO+NNLL) become smaller than those at NLO(NNLO) beyond
$Q=700(400)$ GeV. While for NNLO, the scale uncertainties reach up to $4.5\%$ in high Q regions, the corresponding ones at NNLO+NNLL level reach up to $3.2\%$.

In the low $Q$ region, the resummation is found to enhance the scale uncertainties compared to the corresponding fixed-order results. 
This is due to the fact that in the low $Q$ region, the regular terms coming from the 
hard non-collinear gluons as well as those from the other parton
channels are also important. However being nonuniversal, they cannot be resummed to all orders in the perturbation theory. As a result, the resummation cannot improve the scale uncertainties in the fixed-order results in the low $Q$
region.

In the left panel of \fig{fig:match_ZZ_mur}, we present the uncertainties due 
to only the renormalization scale by keeping the $\mu_F=Q$ fixed. Similarly,
in the right panel, the uncertainties due to only factorization scale variations for fixed $\mu_R=Q$ are given. It is also worth noting that while performing the resummation, the large partonic threshold logarithms are resummed to all orders in the perturbation series that is expanded in $\as$, and hence the uncertainties due to $\mu_R$ are expected to be smaller for a given $\mu_F$ as shown in the left panel of \fig{fig:match_ZZ_mur}. However, the scale $\mu_F$ enters both the PDFs as well as the parton coefficient functions. The uncertainty due to $\mu_F$ need not decrease as a result of resummation where the PDFs used are extracted at a particular fixed-order. Such a behavior of $\mu_F$ scale uncertainty can be seen in the right panel of \fig{fig:match_ZZ_mur} and has already been reported in the literature \cite{Banerjee:2017cfc,Banerjee:2018vvb,AH:2019phz,Das:2019bxi,Das:2020gie,AH:2020cok,AH:2020iki,AH:2021kvg,AH:2021vdc,Bhattacharya:2021hae,Das:2022zie,Ravindran:2022aqr,Das:2024auk}.

To study the convergence of the perturbation series, it is useful to keep the PDFs fixed and study 
how the cross sections vary at different orders. For this, we present the K-factors obtained from the invariant mass distribution
computed with NNLO PDFs, to NNLO+NNLL accuracy. Thus, the same $\as$ is used both at NLO and NNLO.
In the top left panel of \fig{fig:match_ZZ_kfacnnlo}, we present the fixed-order K-factors, 
$K_{10}$, $K_{20}$ and $K_{21}$. For a faster converging perturbation 
series, the factor $K_{i,i-1}$ is supposed to be as close to unity as possible. While the $K_{10}$ has the usual
NLO K-factor information and is as large as  $55\% $, from the $K_{21}$ we see that the NNLO corrections could
contribute an additional $25\%$ of NLO results. From the behavior of $K_{21}$, we see that the second-order corrections are
small, but they increase with $Q$. Similar results are presented but for $L_{ij}$ in the top right panel of \fig{fig:match_ZZ_kfacnnlo}. By definition, these ratios will estimate the contribution of higher-order corrections over and above at least LO+LL level. Hence, these are 
smaller than the corresponding fixed-order K-factors $K_{ij}$. 
In the bottom left panel of \fig{fig:match_ZZ_kfacnnlo}, we present the resummed K-factors $R_{i0}$ to estimate the size of the resummed results above the LO predictions.
We notice that the difference ($R_{20} - R_{10}$) is less than ($R_{10} - R_{00}$) for the whole invariant mass region considered here.
In the bottom right panel of \fig{fig:match_ZZ_kfacnnlo}, we plot $R_{ii}$ that gives the information about the resummed contributions over and above the corresponding 
fixed-order corrections. We notice that the contribution from the resummed corrections for any given $i$ is smooth but slowly increasing 
as $Q$ increases. From $R_{22}$, we can estimate the size of higher logarithmic terms beyond NNLO to be around $2.8\%$ of NNLO for  $Q=1300$ GeV.


The total production cross sections, after integrating over the invariant mass region over the full kinematic region,
for LHC energies are substantially large for the $ Z$-boson pair production process. These production cross sections
have been given in the \tab{tab:tableZZ} up to NNLO+NNLL accuracy, along with the theory uncertainties due to the seven-point
scale variations for different centre of mass energies of the incoming protons. We notice that for any given
centre of mass energy, the cross sections increase while the uncertainties decrease as we go from LO to NLO,
in the fixed-order case. However, the uncertainties will increase from NLO to NNLO due to the gluon fusion channel
opening up from the second-order in perturbation theory, and hence new contributions will add up to the renormalization
scale uncertainties.  The gluon fluxes for the LHC energies near the $2 \mz$ 
region are quite large, and this gluon fusion channel contributes about $9.3\%$ of LO $Z$-boson pair production at $13.6$ TeV LHC energy.
To systematically quantify the uncertainties in the perturbation theory, we define the second-order cross section
without this gluon fusion channel and call it NNLO$_{q\bar{q}}$. The scale uncertainties in the total production cross sections, LO, NLO and NNLO$_{q\bar{q}}$ are found to systematically decrease from $4.24\%$ to $1.07\%$ for 
$13.6$ TeV LHC energy. This behavior remains similar for other centre of mass energies. We notice a similar 
behavior in the total production cross sections after the resummation has been performed, i.e. the scale
uncertainties decrease from  $4.62\%$ to $1.42\%$, as we go from LO+LL to NNLO+NNLL. However, the NNLO$_{q\bar{q}}$+NNLL has a somewhat larger scale
uncertainty compared to the one in NNLO$_{q\bar{q}}$. This is simply because of the definition of total production cross
section where the invariant mass has been integrated out from $Q_{\rm min} = 2 \mz$ to $\sqrt{S}$. In the lower 
$Q$-region the contribution from other channels like $qg$-subprocess cannot be ignored. However, with increasing
$Q_{\rm min}$ in the total production cross section, the scale uncertainties in NNLO$_{q\bar{q}}$+NNLL are expected to be smaller
than those in NNLO$_{q\bar{q}}$, as evident from \fig{fig:match_ZZ_mu}.
The NLO corrections for the gluon fusion channel, in the massless quark limit, are about $68\%$ of its LO for the current LHC energies \cite{Caola:2015psa}. The inclusion of massive top quark loops is found to increase this correction to about $73\%$ \cite{Agarwal:2024pod}.
Finally, at this NNLO+NNLL accuracy in the perturbation theory, the NLO EW correction for $Z$-boson pair production also becomes important. The EW corrections are negative, and for total production, they 
are found to be around $-6.2\%$, while the mixed NNLO QCD$\times$EW corrections are about $-5.7\% $
for $13$ TeV LHC. For the invariant mass distribution, the EW corrections amount to $-15\%$  at $1$ TeV \cite{Grazzini:2019jkl}.
Finally,
the NNLO$+$NNLL results match well within the theoretical and experimental uncertainties to the ATLAS \cite{ATLAS:2017bcd} and CMS \cite{CMS:2017dzg} measurements of the $Z$-boson pair production cross section at $13$ TeV. 
The NNLO$+$NNLL enhances the NNLO cross section by $0.6 \%$ of NNLO at $13$ TeV.
We also note that the uncertainties due to the non-perturbative PDFs are also important. These PDF uncertainties for the $q\bar{q}$ initiated
DY processes are found to be about $3\%$ around the 1 TeV region \cite{Das:2022zie}. The $ZZ$ production process also takes place via $q\bar{q}$ initial states,
and hence a similar PDF uncertainty is expected around $Q =1$ TeV.

\input{tableZZ}


%% file: tableZZ.tex
\begin{table}[ht!]
	\begin{center}
{\scriptsize		
\resizebox{15.0cm}{2.8cm}{
		\begin{tabular}{|c|c|c|c|}
\hline
			$\sqrt{S}$  &  $13.0$ TeV & 13.6 TeV& 
			$100.0$ TeV \\

\hline	
\hline
LO                       &  $10.958 \pm 4.00\%$ pb &  $11.664 \pm 4.24\%$ pb & 
			$138.617 \pm 13.84\%$ pb\\
\hline
NLO                      &  $ 14.380 \pm 1.92\%$ pb  &  $ 15.284 \pm 1.91\%$ pb  &
			$169.090 \pm 5.13\%$ pb\\
\hline
			NNLO$_{q\bar{q}}$                     &  $15.427 \pm 1.01\%$ pb  &  $16.437 \pm 1.07\%$ pb  &
			$184.438 \pm 1.69\%$ pb\\
\hline
			NNLO                     &  $16.418 \pm 2.22\%$ pb  &  $17.521 \pm 2.30\%$ pb  &
			$212.344 \pm 3.81\%$ pb\\
\hline
			LO+LL                    &  $12.353 \pm 4.37\%$ pb  &  $13.143 \pm 4.62\%$ pb  & 
			$153.653 \pm 14.10\%$ pb\\
\hline
			NLO+NLL                  &  $14.890 \pm 4.49\%$ pb  &  $15.824 \pm 4.56\%$ pb  &
			$174.162 \pm 7.76\%$ pb\\
\hline
			NNLO$_{q\bar{q}}$+NNLL                &  $15.527 \pm 1.39\%$ pb  &  $16.543 \pm 1.42\%$ pb  &
			$185.348 \pm 2.53\%$ pb\\
\hline
			NNLO+NNLL                &  $16.518 \pm 1.84\%$ pb  &  $17.627 \pm 1.93\%$ pb  &
			$213.254 \pm 3.64\%$ pb\\
\hline
\hline 
ATLAS  ($13.0$ TeV) ~\cite{ATLAS:2017bcd} & \multicolumn{3}{c|}{  $17.3 \pm 0.9$ pb } \\ \hline
CMS  ($13.0$ TeV) ~\cite{CMS:2017dzg} &  \multicolumn{3}{c|}{ $17.2 \pm 1.0$ pb } \\
\hline
\end{tabular}
 }
		\caption{\small{Inclusive cross section for $Z$-boson pair production for different center of mass energies of the incoming protons, along with the corresponding 7-point scale uncertainties. Also, the ATLAS \cite{ATLAS:2017bcd} and CMS\cite{CMS:2017dzg} measurements with the total uncertainty from statistical, systematic and luminosity uncertainties involved in the measurement.
		} 
\label{tab:tableZZ}
 }}
 \end{center}
\end{table}

%% file: conclusion.tex
To summarize, we have performed the threshold resummation for the production 
of a pair of $Z$-bosons at the energies of LHC, by resumming the threshold logarithms to NNLL
accuracy in QCD. The final state having two massive particles makes the process 
dependent one-loop and two-loop virtual corrections more difficult compared to the massless final state 
like di-lepton or di-photon, or one-massive final state like Higgs boson. 
The presence of such virtual amplitudes makes the resummation a challenging task to achieve numerically, in contrast to the case of $2 \to 1$ processes like DY and Higgs production cross-section processes. 
It is also worth noting that compared to the previously available NLO+NNLL results, in the present NNLO+NNLL results, the presence of real corrections is also equally important and thanks to MATRIX where such contributions have already been taken into account. 
In this work, we have performed this resummation by systematically matching to the known fixed order NNLO results (from the package ${\tt MATRIX}$) and present our phenomenological results to NNLO+NNLL accuracy for both the total production cross sections as well as for the invariant mass distribution of the $Z$-boson pair, for the current LHC energies, $13.0$ and $13.6$ TeV, as well as for the upcoming future $100$ TeV collider. 
We notice that the NNLL resummed results in general enhance the cross sections and contribute an additional few percent to the known NNLO results.
After performing the resummation at NNLO$+$NNLL accuracy, our results for the total production at $13$ TeV center of mass energy are found to agree with experimental measurements (within the reported errors) from the ATLAS and CMS.
We have also presented the theory uncertainties by varying the unphysical renormalization and factorization scales from $Q/2$ to $2Q$. 
We find that, after performing the resummation, the scale uncertainties of about $4.5\%$ in NNLO cross
sections get reduced to about $3.2\%$ at NNLO+NNLL level for the invariant mass region $Q=1.3$ TeV.

The availability of these resummed results is expected to provide a foundation for future theoretical and experimental studies of diboson processes at the LHC and beyond.

%% file: appendix.tex
\appendix
\section{Resummation coefficients}
\label{appendixa}
The process-dependent $g_0$ coefficients defined in \eq{eq:g0} are given as 
(defining $L_{qr} = \ln \left(Q^2/\mur^2 \right), L_{fr} = \ln \left( \muf^2/\mur^2\right)$),

\begin{align} 
	\begin{autobreak} 
		g_{01} = 
		2 \mathcal{K}_{(0,1)} 
		+ 2  \Cf \bigg\{ 
		- 3 \Lfr 
		+ \Lqr^2
		+ \z2
		\bigg\},
	\end{autobreak} 
	\\ 
	\begin{autobreak} 
		g_{02} = 
		\mathcal{K}_{(1,1)} 
		+ 2 \mathcal{K}_{(0,2)}
		+ \Cf \nf \bigg\{ 		
		- \frac{328}{21}
		+ \frac{2}{3}\Lfr
		+ \frac{112}{27}\Lqr
		- \frac{20}{9}\Lqr^2
		- \frac{10}{9}\z{2}
		+ \frac{16}{3}\Lfr \z{2}
		- \frac{4}{3}\Lqr \z{2}
		+ \frac{32}{3}\z{3}
		\bigg\}
		+ \Cf^2 \bigg\{ 
		- 3 \Lfr
		+ 18 \Lfr^2
		- 12 \Lfr \Lqr^2
		+ 2 \Lqr^4 
		+ 12 \Lfr \z{2}
		+ 4 \Lqr^2 \z{2}
		+ 2 \z{2}^2
		- 48 \Lfr \z{3}
		\bigg\}
		+ \Cf \bigg\{
		+ 3 \beta_{0} \Lfr^2
		- \frac{2}{3} \beta_{0} \Lqr^3
		-12 \Lfr \mathcal{K}_{(0,1)}
		+ 4 \Lqr^2 \mathcal{K}_{(0,1)}
		- 2 \beta_{0} \Lqr \z{2}
		+ 4 \mathcal{K}_{(0,1)} \z{2}
		+ \frac{46}{3}\beta_{0} \z{3}
		\bigg\}
		+\Cf \Ca \biggl\{
		+ \frac{2428}{81}
		- \frac{17}{3} \Lfr
		-\frac{808}{27} \Lqr
		+ \frac{134}{9} \Lqr^2
		+ \frac{67}{9} \z{2}
		- \frac{88}{3}\Lfr \z{2}
		+ \frac{22}{3} \Lqr \z{2}
		- 4 \Lqr^2 \z{2}
		- 12 \z{2}^2
		-\frac{176}{3} \z{3}
		+ 24 \Lfr \z{3}
		+ 28 \Lqr \z{3}
		\biggr\}
	\end{autobreak} 
\end{align}

Here, we define
\begin{align}
    \begin{autobreak}
	    \mathcal{K}_{(m,n)} = \frac{\mathcal{M}_{(m,n)}^{\text{fin}}}{\mathcal{M}_{(0,0)}}
    \end{autobreak}
\end{align}
where,
\begin{align}
    \begin{autobreak}
	    \mathcal{M}_{(m,n)}^{\text{fin}} \equiv \langle \mathcal{M}_{m}  \rvert \mathcal{M}_{n} \rangle
    \end{autobreak}
\end{align}
and $ \rvert \mathcal{M}_{n} \rangle$ represents the UV-renormalized, IR-finite virtual amplitude at the n-th order in $\as$, as given in Eq. (C3) of Ref. \cite{Ahmed:2020nci}.
The one loop virtual contribution for $ Z$-boson pair production is,

\begin{align}
	\begin{autobreak}
		\mathcal{M}_{(0,1)} = \as(\mu_R^2)\frac{\Gamma(1-\epsilon)}{\Gamma(1-2 \epsilon)}(4 \pi)^\epsilon
		  \bigg(
		  \frac{\mu_R^2}{s}
		  \bigg)^{\epsilon}
		  \cf
		  \biggl[
		- 4
		  \biggl\{
		  \frac{1}{\epsilon^2}
		+ \frac{3}{2 \epsilon}
		  \biggr\}
		   \mathcal{M}_{(0,0)}
		+ \mathcal{M}_{(0,1)}^{f}
		 \biggr].
	\end{autobreak}
\end{align}
here,
\begin{align*}
	\begin{autobreak}
		\mathcal{M}_{(0,1)}^{f} =  B_{f} {\rm N}
	    ( c_v^4 + 6c_v^2c_a^2 + c_a^4 )
	    \bigg( \frac{4}{t^2u^2} \bigg)
	     \bigg[
	     2 \z{2}
	     \big(
	    - 20\mz^2 tu(t + u)
	    - 4\mz^4 (t^2 - 8tu + u^2)
	    + tu(5t^2 + 4tu + 5u^2)
	    + \frac{tu  }{\kappa  s (4\mz^2 - s)^2}
	     \big(
	    - 32\mz^6 tu
	    - 64\mz^8 (t + u)
	    + 8\mz^2 tu (t + u)^2
	    - (t + u)^3 (3t^2 + 4tu + 3u^2)
	    + \mz^4 (22t^3 + 82t^2 u + 82tu^2 + 22u^3)
	    \big)
	    \big)
	    +  {\rm ln}\biggl(\frac{s}{\mz^2}\biggr) \frac{1}{(4\mz^2 - s)^2}
	    \big(
	     8\mz^2 t^2 u^2 (t + u)
	    + 12\mz^8 (t^2 - 8tu + u^2)
	    - tu(t + u)^2 (3t^2 + 4tu + 3u^2)
	    + 4\mz^6 (3t^3 - 5t^2 u - 5tu^2 + 3u^3)
	    + \mz^4 (3t^4 + 14t^3 u + 78t^2 u^2 + 14tu^3 + 3u^4)
	    \big)
		+  \frac{ t u}{\kappa  s (4\mz^2 - s)^2}\big({\rm ln}(x) + 4 {\rm Li}_{2}(-x) \big)
	    \big(
	    - 32\mz^6 tu
	    - 64\mz^8 (t + u)
	    + 8\mz^2 tu (t + u)^2
	    - (t + u)^3 (3t^2 + 4tu + 3u^2)
	    + \mz^4 (22t^3 + 82t^2 u + 82tu^2 + 22u^3)
	    \big)
	    + \frac{1}{(t - \mz^2)(u - \mz^2)(4\mz^2 - s)}
	    \big(
	     18\mz^{10} (t^2 - 8tu + u^2)
	    + \mz^8 (-9t^3 + 131t^2 u + 131tu^2 - 9u^3)
	    - 7t^2 u^2 (t^3 + t^2 u + tu^2 + u^3)
	    - 4\mz^4 tu (4t^3 + 23t^2 u + 23tu^2 + 4u^3)
	    + \mz^6 (-9t^4 + 14t^3 u - 66t^2 u^2)
	    + 14tu^3
	    - 9u^4
	    \big)
	    + \mz^2 tu (9t^4 + 32t^3 u + 82t^2 u^2 + 32tu^3 + 9u^4)
	    +
	     {\rm ln}\bigg(\frac{-t}{\mz^2}\bigg) \frac{u }{(t - \mz^2)^2}
	    \big(
	    3\mz^8 (-4t + u)
	    + 6\mz^6 t(2t + u)
	    - 2\mz^2 t^2 u(4t + u)
	    + 3t^3 u(2t + 3u)
	    + \mz^4 t(-2t^2 + tu - 3u^2)
	    \big)
	    + \bigg(
	     2 {\rm ln}\bigg(\frac{-t}{\mz^2}\bigg){\rm ln}\bigg( \frac{t-\mz^2}{\mz^2 s}\bigg)
	    + 4 {\rm Li}_{2}\bigg(\frac{t}{\mz^2}\bigg)
	    - {\rm ln}^2\bigg( \frac{-t}{\mz^2} \bigg) \bigg) u
	    \big(
	    \mz^4 (8t - 2u)
	    - 4\mz^2 t(2t + u)
	    + t(2t^2 + 2tu + u^2)
	    \big)

	    + {\rm ln}\bigg(\frac{-u}{\mz^2}\bigg) \frac{t }{(u - \mz^2)^2}
	    \big(
	    3\mz^8 (-4u + t)
	    + 6\mz^6 u(2u + t)
	    - 2\mz^2 u^2 t(4u + t)
	    + 3u^3 t(2u + 3t)
	    + \mz^4 u(-2u^2 + tu - 3t^2)
	    \big)
	    + \bigg(
	    2 {\rm ln}\bigg(\frac{-u}{\mz^2}\bigg){\rm ln}\bigg( \frac{u-\mz^2}{\mz^2 s}\bigg)
	    + 4 {\rm Li}_{2}\bigg(\frac{u}{\mz^2}\bigg)
	    - {\rm ln}^2\bigg( \frac{-u}{\mz^2} \bigg) \bigg) t
	    \big(
	    \mz^4 (8u - 2t)
	    - 4\mz^2 u(2u + t)
	    + u(2u^2 + 2tu + t^2)
	    \big)
	    \bigg] .
	\end{autobreak}

\end{align*}

\begin{align}
	x = \frac{1-\kappa}{1+\kappa} ; \quad
	\kappa = \sqrt{1- 4\frac{m_z^2}{s}}
\end{align}

The process-independent universal resum exponent defined in \eq{eq:gn} which 
are used for DY-type processes are given as,
\begin{align} 
	\begin{autobreak} 
		\gNB1 =
		\bigg[ \AAo  ~  \bigg\{ 2
		- 2 ~ \LogmW1
		+ 2 ~ \LogmW1 ~ \iW \bigg\}
		\bigg],   
	\end{autobreak} 
	\\ 
	\begin{autobreak} 
		\gNB2 =
		\bigg[ \DDo  ~  \bigg\{ \frac{1}{2} ~ \LogmW1 \bigg\}      
		+ \AAt  ~  \bigg\{ 
		- \LogmW1
		- \w \bigg\}      
		+ \AAo  ~  \bigg\{ \bigg(\LogmW1
		+ \frac{1}{2} ~ \LogmW1^2
		+ \w\bigg) ~ \bigg(\frac{\beta_{1}}{\beta_0^{2}}\bigg)
		+ \bigg(\w\bigg) ~ \Lfr
		+ \bigg(\LogmW1\bigg) ~ \Lqr \bigg\}
		\bigg],   
	\end{autobreak} 
	\\ 
	\begin{autobreak} 
		\gNB3 =
		\bigg[ \btzAIII  ~  \bigg\{ 
		- \WbimW
		+ \w \bigg\}      
		+ \btzAII  ~  \bigg\{ \bigg(2 ~ \WbimW\bigg) ~ \Lqr
		+ \bigg(3 ~ \WbimW
		+ 2 ~ \LogomWtIMW
		- \w\bigg) ~ \bigg(\frac{\beta_{1}}{\beta_0^{2}}\bigg)
		+ \bigg(
		- 2 ~ \w\bigg) ~  \Lfr \bigg\}      
		+ \btzAI  ~  \bigg\{ 
		- 4 ~ \z2 ~ \WbimW
		+ \bigg(
		- \LogtmWtIMW
		- \WbimW
		- 2 ~ \LogomWtIMW
		+ 2 ~  \LogmW1
		+ \w\bigg) ~ \bigg(\frac{\beta_{1}}{\beta_0^{2}}\bigg)^2
		+ \bigg(
		- \WbimW\bigg) ~ \Lqr^2
		+ \bigg(
		- \WbimW
		- 2 ~ \LogmW1
		- \w\bigg) ~ \btoo
		+ \bigg(\bigg(
		- 2 ~ \WbimW
		- 2 ~ \LogomWtIMW\bigg) ~ \bigg(\frac{\beta_{1}}{\beta_0^{2}}\bigg)\bigg) ~ \Lqr
		+ \bigg(\w\bigg) ~ \Lfr^2 \bigg\}      
		+ \btzDII  ~  \bigg\{ \WbimW \bigg\}      
		+ \btzDI  ~  \bigg\{ \bigg(
		- \WbimW\bigg) ~ \Lqr
		+ \bigg(
		- \WbimW
		- \LogomWtIMW\bigg) ~ \bigg(\frac{\beta_{1}}{\beta_0^{2}}\bigg) \bigg\}
		\bigg],   
	\end{autobreak} 
\end{align}
Here $A_i$ are the universal cusp anomalous dimensions, $D_i$ are the threshold 
non-cusp anomalous dimensions, and $\omega= 2 \as \beta_0 \ln \overbar{N} $.
Note that all the perturbative quantities are expanded in powers of $\as$. 
The cusp anomalous dimensions $A_i$ are given as (recently four-loops results are available can be found in Ref. \cite{Henn:2019swt, Huber:2019fxe, vonManteuffel:2020vjv}),
\begin{align} 
	\begin{autobreak} 
		\A1 = C_{F}
		\bigg\{ 4
		\bigg\},   
	\end{autobreak} 
	\\ 
	\begin{autobreak} 
		\A2 = C_{F}
		\bigg\{ \nf    \bigg( 
		- \frac{40}{9} \bigg)      
		+ \Ca    \bigg( \frac{268}{9}
		- 8  \z2 \bigg)
		\bigg\},   
	\end{autobreak} 
	\\ 
	\begin{autobreak} 
		\A3 = C_{F}
		\bigg\{ \nf^2    \bigg( 
		- \frac{16}{27} \bigg)      
		+ \Cf  \nf    \bigg( 
		- \frac{110}{3}
		+ 32  \z3 \bigg)      
		+ \Ca  \nf    \bigg( 
		- \frac{836}{27}
		- \frac{112}{3}  \z3
		+ \frac{160}{9}  \z2 \bigg)      
		+ \Ca^2    \bigg( \frac{490}{3}
		+ \frac{88}{3}  \z3
		- \frac{1072}{9}  \z2
		+ \frac{176}{5}  \z2^2 \bigg)
		\bigg\}, 
	\end{autobreak} 
\end{align}

The coefficients $D_i$ are given as,
\begin{align} 
	\begin{autobreak} 
		D_1 = C_{F}
		\bigg\{0
		\bigg\},   
	\end{autobreak} 
	\\ 
	\begin{autobreak} 
		D_2 = C_{F}
		\bigg\{ \nf    \bigg( \frac{224}{27}
		- \frac{32}{3}  \z2 \bigg)      
		+ \Ca    \bigg( 
		- \frac{1616}{27}
		+ 56  \z3
		+ \frac{176}{3}  \z2 \bigg)
		\bigg\},   
	\end{autobreak} 
\end{align}

Here,
\begin{align}
	&C_{A} = {\rm N} \,,~~~
	C_{F}  = \frac{{\rm N}^2-1}{2{\rm N}}  \,, \\
	&\beta_{0} = \frac{11}{3}\Ca - \frac{2}{3} \nf \,,  \\
	&\beta_{1} = \frac{34}{3}\Ca^{2} - \frac{10}{3} \nf \Ca - 2 \nf \Cf ,\\
	&\beta_{2} = \frac{2857}{54} \Ca^3 - \frac{1415}{54} \nf \Ca^2 - \frac{205}{18} \nf \Cf \Ca + \nf \Cf^2 
	             + \frac{79}{54} \nf^2 \Ca + \frac{11}{9} \nf^2 \Cf .
\end{align}